\documentclass[twocolumn,showpacs,superscriptaddress,amsmath]{revtex4}
\usepackage{threeparttable}

\usepackage{graphicx}% Include figure files
\usepackage{dcolumn}% Align table columns on decimal point
\usepackage{bm}% bold math
\usepackage{color}
\usepackage{ifthen}

\newcommand{\mPnwup}{P_5 \begin{pmatrix} a+i&\\a&b\\c&d \end{pmatrix}}
\newcommand{\mPnwle}{P_5 \begin{pmatrix} a+i&a&b\\&c&d \end{pmatrix}}
\newcommand{\mPneup}{P_5 \begin{pmatrix} &b+i\\a&b\\c&d \end{pmatrix}}
\newcommand{\mPneri}{P_5 \begin{pmatrix} a&b&b+i\\c&d& \end{pmatrix}}
\newcommand{\mPsedo}{P_5 \begin{pmatrix} a&b\\c&d\\&d+i \end{pmatrix}}
\newcommand{\mPseri}{P_5 \begin{pmatrix} a&b&\\c&d&d+i \end{pmatrix}}
\newcommand{\mPswdo}{P_5 \begin{pmatrix} a&b\\c&d\\c+i& \end{pmatrix}}
\newcommand{\mPswle}{P_5 \begin{pmatrix} &a&b\\c+i&c&d \end{pmatrix}}

\newcommand{\Pnwup}{P_5 \begin{pmatrix}a&\\a+i&b\\c&d\end{pmatrix}}
\newcommand{\Pnwle}{P_5 \begin{pmatrix}a&a+i&b\\&c&d\end{pmatrix}}
\newcommand{\Pneup}{P_5 \begin{pmatrix}&b\\a&b+i\\c&d\end{pmatrix}}
\newcommand{\Pneri}{P_5 \begin{pmatrix}a&b+i&b\\c&d&\end{pmatrix}}
\newcommand{\Psedo}{P_5 \begin{pmatrix}a&b\\c&d+i\\&d\end{pmatrix}}
\newcommand{\Pseri}{P_5 \begin{pmatrix}a&b&\\c&d+i&d\end{pmatrix}}
\newcommand{\Pswdo}{P_5 \begin{pmatrix}a&b\\c+i&d\\c&\end{pmatrix}}
\newcommand{\Pswle}{P_5 \begin{pmatrix}&a&b\\c&c+i&d\end{pmatrix}}
\newcommand{\fPp}[1]{
\ifthenelse{\equal{#1}{1}}{P_4\begin{pmatrix}a+i&b\\c&d\end{pmatrix}}{}
\ifthenelse{\equal{#1}{2}}{P_4\begin{pmatrix}a&b+i\\c&d\end{pmatrix}}{}
\ifthenelse{\equal{#1}{3}}{P_4\begin{pmatrix}a&b\\c+i&d\end{pmatrix}}{}
\ifthenelse{\equal{#1}{4}}{P_4\begin{pmatrix}a&b\\c&d+i\end{pmatrix}}{}
}
\newcommand{\fPC}{P_4\begin{pmatrix} a&b\\c&d\end{pmatrix}}
\begin{document}
\title{Reentrant phase transition in a predator prey model}
\author{Sung-Guk Han}
\affiliation{Department of Physics and BK21 Physics Research Division, Sungkyunkwan University, Suwon 440-746, Korea}
\author{Su-Chan Park}
\affiliation{Institut f{\"u}r Theoretische Physik, Universit{\"a}t zu 
K{\"o}ln, Z{\"u}lpicher Strasse 77,
	50937 K{\"o}ln, Germany}
\author{Beom Jun Kim}
\email[Corresponding author: ]{beomjun@skku.edu}
\affiliation{BK21 Physics Research Division and Department of Energy Science,
Sungkyunkwan University, Suwon 440-746, Korea}

\begin{abstract}
We numerically investigate the six-species predator-prey game in complex
networks as well as in $d$-dimensional regular 
hypercubic lattices with $d=1,2,\cdots, 6$.  
The food-web topology of the six species contains two directed loops, each of which is composed of
cyclically predating three species. As the mutation rate is lowered below
the well-defined phase transition point, the $Z_2$ symmetry related with the
interchange of the two loops is spontaneously broken, and it has been known
that the system develops the defensive alliance in which three cyclically
predating species defend each other against the invasion of other species. In
the Watts-Strogatz small-world network structure 
characterized by the rewiring probability
$\alpha$, the phase diagram shows the reentrant behavior as $\alpha$ is varied,
indicating a twofold role of the shortcuts. In $d$-dimensional regular
hypercubic lattices, the system also exhibits the reentrant phase transition as
$d$ is increased. We identify universality class of the phase transition
and discuss the proper mean-field limit of the system.
\end{abstract}

\pacs{89.75.Hc, 89.75.Fb, 68.35.Rh, 87.23.Cc}
% 89.75.Hc : Networks and genealogical trees
% 89.75.Fb : Structures and Organization in complex systems 
% 68.35.Rh : Phase transition and critical phenomena
% 87.23.Cc : Population dynamics and ecological pattern formation 
\maketitle

\section{\label{sec:intro}Introduction}

Recently, nonequilibrium dynamical systems as well as 
standard equilibrium statistical mechanical model systems have been intensively 
studied in the interaction 
structure of complex networks~\cite{review}.
Effects of shortcuts in the Watts-Strogatz (WS) networks~\cite{WS} 
on collective dynamic behaviors have been drawn much attention, 
revealing the close interplay between the structural and the dynamical
properties. For networks of highly heterogeneous degree distributions
such as the Barab\'{a}si-Albert (BA) scale-free network~\cite{BA}, 
various aspects of collective behaviors have also been studied.
Lots of existing studies have been performed in the framework
of the phase transition and critical 
behavior~\cite{review}, and 
complex network topology in many cases gives rise to the mean-field 
universality class. 
It has
been agreed that the shortcuts and the hub vertices
play important roles, increasing the effective dimensionality
of the WS and the BA networks, respectively, yielding
the critical behavior beyond upper critical dimension.
In contrast, although various nonequilibrium models 
like game theoretic models,  epidemic spread models, the voter model, 
and the contact process, have been actively studied in the complex network 
research area~\cite{review}, 
generic understanding of how the underlying network topology affects
the nonequilibrium phase transition is still lacking.

In population genetics, the so-called Lotka-Volterra model
has often been studied. In the point of view of statistical physics,
the neglect of the spatial density fluctuation in the original
Lotka-Volterra model corresponds to  a mean-field approximation, which cannot
be justified in general, since every living organism inevitably
lives in a finite-dimensional space with a limited range of
interactions. Accordingly, population genetics models allowing
spatial inhomogeneity in species densities are desirable, which can be
simply realized by putting predators and preys on regular
lattices in two- and three-dimensions (2D and 3D).
Indeed, a cyclically interacting three-species predator-prey model has been
studied both experimentally and numerically in Ref.~\onlinecite{B.Kerr} 
to uncover how the biodiversity of the system can be maintained.
Similar three-species predator-prey model has also been investigated
not in a regular lattice structure but in the small-world interaction 
structure~\cite{G.Szabo2}.
More complicated predator-prey model  has been suggested 
later~\cite{G.Szabo1,G.Szabo3,B.J.Kim2}, with six
or nine species interacting with each other in a given
food-web structure. In these studies it has been found that
a subgroup of cyclically predating species is spontaneously formed 
and species within the subgroup protect the member species
from the attacks by other species outside of the subgroup.
The spontaneous formation of such a defensive alliance is well
captured by the statistical mechanical approach and the research
focus has been put on the existence and the nature of the 
phase transition of the spontaneous formation of alliance as the
mutation rate is varied. 

\begin{figure}
\includegraphics[width=0.42\textwidth]{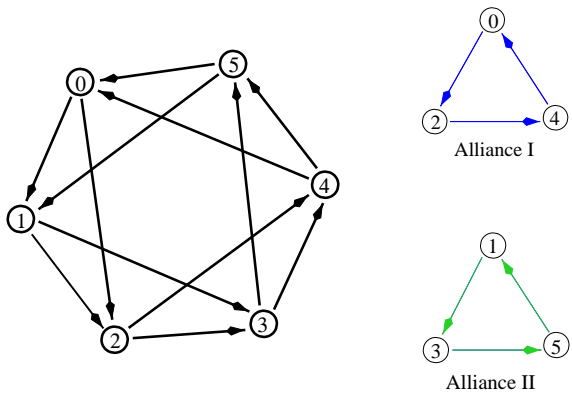}
\caption{\label{fig:foodweb}
(Color online) Food web of the six-species predator-prey model. 
Each species has two predators, two preys, and two noninteracting
neutral species. The two groups, called defensive alliances
(alliance I and II), can be formed.
}
\end{figure}

In the present work, we study the six-species predator-prey game
on various spatial interaction structures, in the context of
the phase transition and the critical behavior of a nonequilibrium statistical
physics model in complex networks. The food web under
consideration (shown in Fig.~\ref{fig:foodweb})
has first been introduced in Ref.~\onlinecite{G.Szabo3}
and later studied in Ref.~\onlinecite{B.J.Kim2}.
Specifically, Szab{\'o} {\it et al.}~\cite{G.Szabo3} have shown
that the predator-prey game played on a 2D regular lattice
exhibits a phase transition of the 2D Ising universality as the mutation
rate is changed, and also identified
the relevant order parameter detecting the $Z_2$ symmetry breaking
of the defensive alliance of three cyclically predating species (see
Fig.~\ref{fig:foodweb}). In what follows, we will refer to this model as 
the defensive alliance process (DAP).
In Ref.~\onlinecite{B.J.Kim2} the DAP
has been studied on the WS network structure and the
effects played by the two different types of randomness, i.e.,  
the temporal randomness induced by the mutation, and the structural
randomness introduced by the shortcuts in the WS network, have been
investigated. It has also been observed that the DAP
in the WS network shows a discontinuous phase transition at 
nonzero rewiring probability different from the continuous transition 
in 2D regular lattice~\cite{B.J.Kim2}.
We  extend in this work the study in Ref.~\onlinecite{B.J.Kim2}
for the WS networks and construct the complete phase diagram in the plane of
the two different randomness. 
To investigate the role of the structural inhomogeneity,
the nature of the phase transition
in the regular $d$-D hypercubic lattices is also
numerically studied. Since the mean-field theory developed in Ref.~\onlinecite{B.J.Kim2}
does not predict a nontrivial critical point, we perform the cluster mean-field (CMF)
approximation for 2D model with the hope for the better understanding as to the nature
of the transition in higher dimensions.

The present paper is organized as follows: Section~\ref{sec:WS} presents
our results of the phase diagram for the DAP
on the WS networks as well as on the $d$-dimensional hypercubic lattices with
$d=1,2,\cdots,6$ and discusses the reentrance transition and the nature
of the phase transition.  In Sec.~\ref{sec:cluster}, the cluster
mean-field theory with the cluster size 2 is developed for the 2D defensive alliance model.
Finally, we summarize our results in Sec.~\ref{sec:summary}.

\section{\label{sec:WS}Defensive alliance process}
We start from the description of how the DAP is implemented on general
networks.
The algorithm for the simulations is as follows:
Initially at time $t=0$, six species are equally distributed on vertices 
of the given network structure, with the density $c_s$ of the species
$s$ given by $c_s(t=0) = 1/6$ for all species ($s=0,1, \cdots, 5)$.  
At each time step, one vertex is chosen at random and 
with the probability $P$ the species at the vertex is mutated 
to one of its predators. Otherwise, with probability $1-P$, the species 
at the vertex plays the predator-prey game according to the rules depicted
in Fig.~\ref{fig:foodweb} with one of its randomly 
selected neighbors, and the winner between the two occupies
the loser's vertex. If the two species are neutral, i.e., if no
arrow connects the two in Fig.~\ref{fig:foodweb}, the game will end in a draw
and nothing happens. The above procedure is repeated until the system
approaches the steady state. 

Only for convenience, we define the parameter $\mu$ as~\cite{B.J.Kim2} 
\begin{equation}
\label{eq:mu}
\mu = \ln(1/P),
\end{equation}
which is a decreasing function of $P$.
As $P$ becomes larger, the temporal randomness becomes stronger. In other words,
$\mu$ resembles an inverse temperature of equilibrium systems
and  $1/\mu$ can be interpreted as an effective temperature.

In order to detect the alliance breaking transition we measure 
the order parameter $m$ (we also call it the magnetization in analogy to
the ferromagnetic Ising model) defined by~\cite{G.Szabo3,B.J.Kim2} 
\begin{equation}
\label{eq:m}
m = \langle | (c_{0}+c_{2}+c_{4})-(c_{1}+c_{3}+c_{5}) |\rangle,
\end{equation}
where $\langle \dots \rangle$ is the time average after the steady
state is achieved.
In this work, a sufficiently long equilibration time (20 000 Monte-Carlo steps)
is taken.  As $P$ becomes larger toward unity, all species are equally probable 
and $m(P \rightarrow 1) = 0$, while as $P$ approaches zero, the
spontaneous development of defensive alliances gives us $m \approx 1$
both for the alliances I and II (see Fig.~\ref{fig:foodweb}), 
indicating the possibility of a phase transition at nontrivial critical point $\mu_c$. 

In simulating DAP, we had a numerical difficulty especially when 
the mutation probability $P$ is very small. In such cases, 
the system often spends very long time before achieving the steady state, 
which we try to avoid through the use of the simulated annealing technique 
in statistical physics by lowering $P$ slowly starting from a 
high value of $P$. 
It is also observed that if the system size 
is not sufficiently large, the population becomes monomorphic and 
all vertices are occupied by a single species before mutation acts. 
The rare mutation can produce a predator of the species,
but the small system at the low mutation probability will again be monomorphic
unless a predator of the new born predator is generated by another mutation. 
The results presented below are for sufficiently large systems where
the polymorphic population is attained.

\subsection{DAP in the WS networks}
\label{subsec:WS}

We first construct WS networks for the DAP as follows~\cite{WS}: 
(i) 1D  and 2D regular lattices are first built. 
For 1D each vertex has connections to its four neighbors (nearest
and next-nearest neighbors), whereas for 2D we assume only 
four nearest neighbor connections. (ii) With the rewiring probability
$\alpha$ one end of each local link is moved to a randomly chosen
other lattice point. Throughout this paper, we call the resulting
networks as WS$_1$ and WS$_2$, respectively, in order to indicate
the dimensionality $1$ and $2$ of {\em initial} regular lattices
from which the WS network is built as described above. 
However, the subscripts $1$ and $2$ in WS$_1$ and WS$_2$
should not be interpreted as the dimensionality of the resulting
small-world networks.  The structure of the network is 
varied with $\alpha$ from a regular network ($\alpha=0$) to a fully random 
network ($\alpha=1$). 

\begin{figure}
\includegraphics[width=0.48\textwidth]{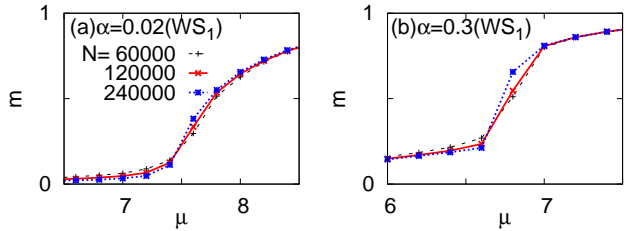}
\caption{
(Color online) 
The order parameter $m$ detecting the alliance breaking transition
versus the mutation parameter $\mu = \ln(1/P)$ with the mutation
probability $P$ is shown for the network WS$_1$, constructed from
the 1D regular lattice via rewiring,  at the
rewiring probability (a) $\alpha = 0.02$ and (b) $\alpha = 0.3$.
As $\mu$ is increased, i.e., as the mutation probability $P$ is decreased,
a spontaneous formation of the defensive alliance occurs.
The transition point $\mu_c$ is roughly estimated from the comparison 
of different sizes ($N=60\;000, 120\;000, 240\;000$): $\mu_c = 7.4 \pm 0.1$
for (a) $\alpha = 0.02$ and $\mu_c = 6.8 \pm 0.1$ for (b) $\alpha = 0.3$.
}
\label{fig:1dmmu}
\end{figure}

\begin{figure}
\includegraphics[width=0.48\textwidth]{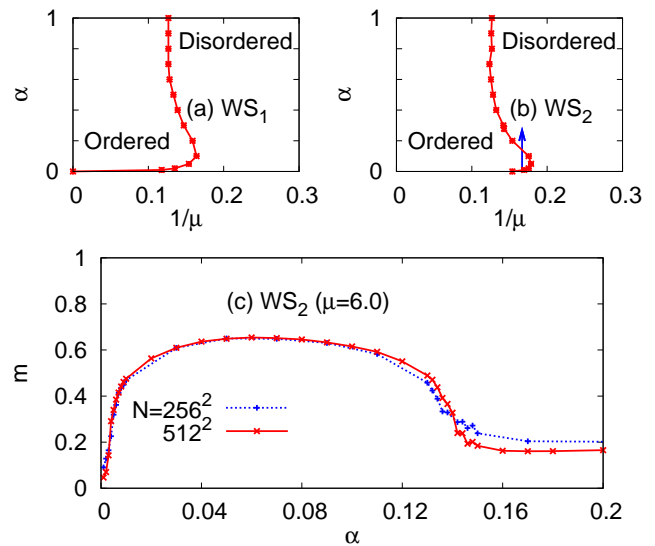}
\caption{\label{fig:wsphd}
(Color online) Phase diagrams for the DAP
in the plane of the mutation parameter $1/\mu [= 1/\ln(1/P)]$ with the mutation 
rate $P$ and the rewiring probability
$\alpha$ for (a) WS$_1$ and (b) WS$_2$ networks.
As the rewiring probability $\alpha$ is changed, both (a) and (b) show
reentrant phase transitions. A discontinuous phase transition
is observed at any nonzero value of $\alpha$. At $\alpha = 0$,
WS$_1$ and WS$_2$ correspond to the regular 1D and 2D lattices.
(c) $m$  versus $\alpha$ for the WS$_2$ networks of the sizes
$N=256 \times 256$ and $512 \times 512$ with $\mu$ set to $6.0$, following
the vertical arrow in (b). The reentrant behavior is again seen very
clearly.
}
\end{figure}

Except for the case of $\alpha = 0$, corresponding to 
locally connected 1D and 2D regular 
lattices, the phase transition is found to be of the discontinuous nature as
was already found in Ref.~\onlinecite{B.J.Kim2}: 
When $\mu \gtrsim \mu_c$, the order parameter $m$ saturates
toward a finite value from below as the system size $N$ is increased, while
at $\mu \lesssim \mu_c$, it decreases with $N$, which implies
that in the thermodynamic limit of $N \rightarrow \infty$, the order
parameter changes abruptly, signalling a discontinuous phase transition
(see Ref.~\onlinecite{B.J.Kim2} for more detailed discussion). 
For examples,
$m$ versus $\mu$ for the WS$_1$ network is shown in Fig.~\ref{fig:1dmmu}
for $\alpha=$ (a) 0.02 and (b) 0.3, with the estimation 
$\mu_c = 7.4(1)$ and $\mu_c = 6.8(1)$, respectively.
We use this finite-size behavior to locate $\mu_{c}$ as above, 
which is then used to construct the full phase diagrams in 
Fig.~\ref{fig:wsphd}. 

It is very interesting that the phase diagrams in Fig.~\ref{fig:wsphd} 
exhibit reentrant behaviors as $\alpha$ is increased. This indicates
that the role of the random shortcuts is somehow twofold,
facilitating the defensive alliance at small $\alpha$ and
then suppressing the formation of the alliance at large $\alpha$.
In detail, Fig.~\ref{fig:wsphd}(c) displays $m$ versus 
the rewiring probability $\alpha$ at 
fixed $\mu=6.0~(1/\mu \approx 0.17)$ for the WS$_2$ network, following
the vertical arrow indicated in Fig.~\ref{fig:wsphd}(b).
It is shown clearly that as $\alpha$ increases the system starts
from a disordered phase with a very small $m$ and enters an ordered
phase and then finally leaves back to a disordered phase.
We also check the finite-size effect by comparing $m$ for two different
sizes $N=256\times 256$ and $512\times 512$ in Fig.~\ref{fig:wsphd}(c);
the existence of ordered phase in the intermediate region of $\alpha$
is shown to be not a finite-size artifact. 
As $\alpha$ is increased from zero,  more shortcuts make the system
more strongly correlated in the sense that the change of the 
dynamic state of one vertex can affect more vertices due to the
small-world effect~\cite{WS}. Consequently, we believe that
the first increase of $m$ for small $\alpha$ can be attributed
to the strengthened correlation due to more shortcuts.
As $\alpha$ is increased further, the reduction of the path lengths
by more shortcuts becomes less influential, and more shortcuts
appear to introduce stronger spatial randomness, which eventually
makes the system less ordered.

In our simulations, we also notice that the behavior of $m$ upon the
change of $\mu$ is different for small $\alpha$ and large $\alpha$:
When $\alpha$ is smaller than some value ($\alpha \lesssim \alpha^*$), 
the order parameter in the disordered phase  $m(\mu \lesssim \mu_c)$
approaches zero as $N$ is increased [see Fig.~\ref{fig:1dmmu}(a)
for the WS$_1$ network at $\alpha = 0.02$]. In comparison,
at $\alpha \gtrsim \alpha^*$, $m(\mu \lesssim \mu_c)$ remains finite
as $N$ becomes larger as one can see in Fig.~\ref{fig:1dmmu}(b)
for $\alpha = 0.3$. It appears that $\alpha^*$
is close to the value of $\alpha$ at the end point of the lob 
structure in the phase diagram, i.e., $\alpha^* \approx 0.1$ for the
WS$_1$ network.

\begin{figure*}
\includegraphics[width=0.96\textwidth]{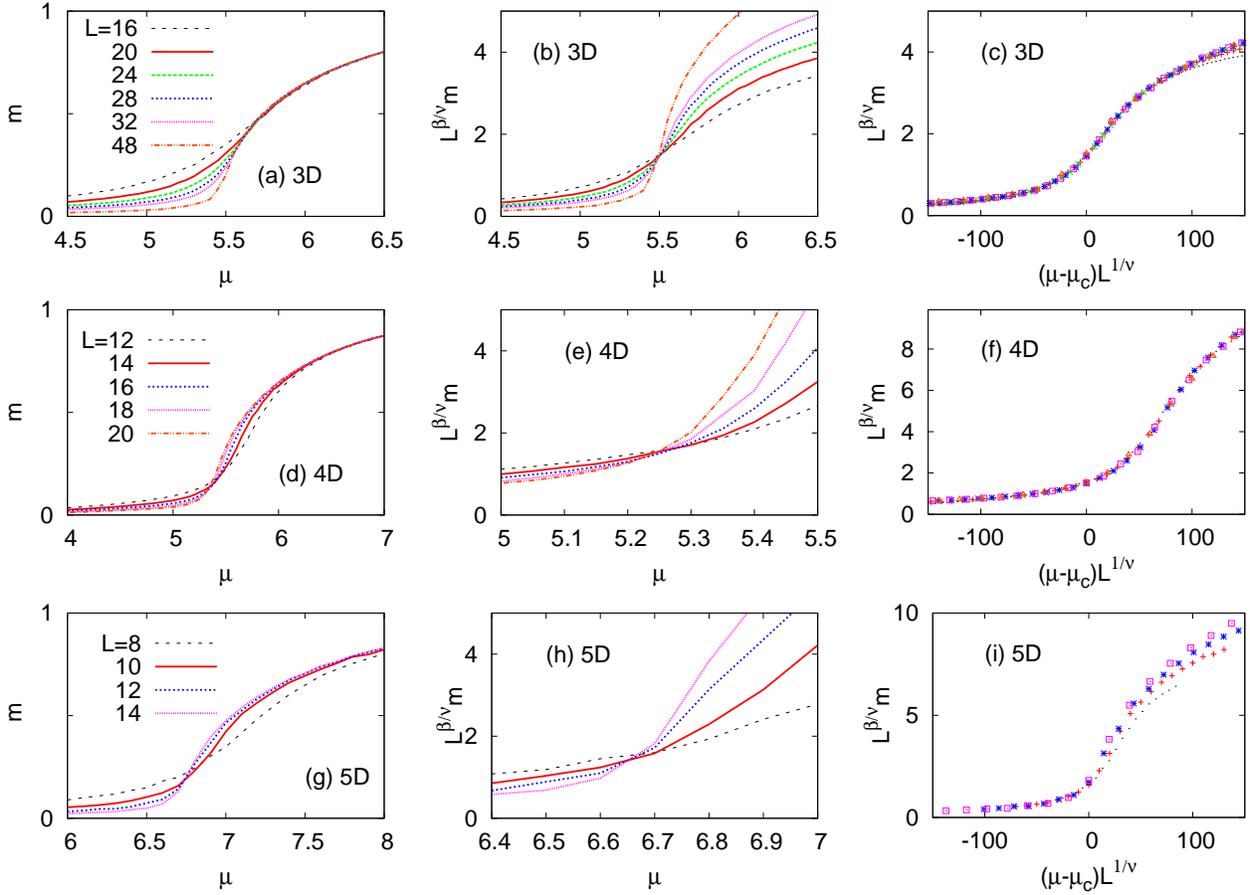}
\caption{(Color online) Phase transitions in 3D [for (a)-(c)], 4D [for (d)-(f)],
and 5D [for (g)-(i)] regular hypercubic lattices in terms of the mutation parameter $\mu$. 
$m$ versus $\mu$, $L^{\beta/\nu}m$ versus $\mu$, and
$L^{\beta/\nu}m$ versus $(\mu - \mu_c)L^{1/\nu}$ are shown
in (a), (b), (c) for 3D, (d), (e), (f) for 4D, and (g), (h), (i) for 5D, 
respectively.
After the determination of the critical point $\mu_c$ in (b), (e), (h),
all data points collapse to single smooth curve by using the
finite-size scaling form in Eq.~(\ref{eq:mfinite}). 
In 3D, $\mu_{c} \approx 5.5$, $\beta \approx 0.33$, and $\nu \approx 0.63$ 
are obtained in accord with the 3D Ising universality class.
In 4D, on the other hand, we obtain 
$\mu_{c} \approx 5.25$, $\beta \approx \nu \approx 0.5$, which are the 
Ising mean-field exponents. In 5D, our simulation results are again consistent
with the mean-field exponents $\beta \approx \nu \approx 0.5$ with $\mu_c \approx 6.7$.
}
\label{fig:345d}
\end{figure*}

At $\alpha = 0$, WS networks do not possess any shortcuts
and thus correspond to locally connected regular lattices.
Observed phase transitions here at $\alpha = 0$ for both WS$_1$ and WS$_2$ 
networks are consistent with the simple expectation that due to the underlying 
$Z_2$ symmetry of the two defensive alliances, the 
DAP should belong to the same universality class
as the equilibrium Ising models in 1D and 2D regular lattices, 
i.e.,  no phase transition at 
finite $\mu$ for 1D and the phase transition with the 2D Ising
critical exponents at finite $\mu_c$ for 2D~\cite{G.Szabo3,B.J.Kim2}. 
However, the existence of discontinuous phase transition at 
nonzero $\alpha$ in WS$_1$ and WS$_2$ networks clearly contradicts 
the above simple naive expectation: It has been
known that standard equilibrium models in statistical mechanics such as
the Ising and the $XY$ models in the WS networks exhibit the mean-field type 
continuous phase transition, which has been
attributed to  the effective increase of the spatial dimensionality
due to the shortcuts (see, e.g.,  Refs.~\onlinecite{Hong_Kim}).
This clearly gives a caveat that one needs to be careful in generalizing  conclusions 
drawn for equilibrium models to nonequilibrium models.

The question we are now addressing is if the structural inhomogeneity 
is responsible for the discontinuous transition. Since the mean-field theory
developed in Ref.~\onlinecite{B.J.Kim2} cannot predict a nontrivial critical point,
the nature of the transition on the WS network cannot be understood from 
this theory.
So it is inevitable to study regular higher-dimensional systems, 
especially beyond the
upper critical dimensions of the Ising class. For completeness,
we will study the DAP in regular 3-, 4-, and 5-dimensional hypercubic lattices
in the next section, before studying 6-dimensional system in Sec.~\ref{subsec:6d}.

\subsection{DAP in 3-, 4-, and 5-dimensional regular lattices}
The results in three-,  four-, and five-dimensional regular hypercubic lattices 
are given in Fig.~\ref{fig:345d}
for hypercubic lattices with the linear size $L$ (the total number of individual species $N = L^d$). 
We use the system size  $L=16, 20, 24, 28, 32$, and $48$ for 3D,  
$L=12, 14, 16, 18$, and 20 for 4D, and $L=8, 10, 12$ and 14 for 5D, 
respectively.
During numerical simulations, we measure
the order parameter $m$ in Eq.~(\ref{eq:m}) and use the standard
finite-size scaling form of the magnetization: 
\begin{equation}
\label{eq:mfinite}
m = L^{-\beta/\nu}f\left ((\mu-\mu_{c})L^{1/\nu}\right ), 
\end{equation}
where $f(x)$ is a suitable scaling function with the scaling
variable $x$, and $\beta$ and $\nu$ are critical exponents for
the order parameter and the correlation length, 
respectively~\cite{N.Goldenfeld}.
Figure~\ref{fig:345d} summarizes the numerical results for 
the phase transition in the 3D [for (a)-(c)], 4D [for (d)-(f)], 
and 5D [for (g)-(i)] regular hypercubic lattices. Clearly exhibited is the vanishing 
of the order parameter at low $\mu$ [see Fig.~\ref{fig:345d} (a), (d), and (g),
in which $m$ versus $\mu$ is shown for 3D, 4D, and 5D].
The critical point $\mu_c$ is determined from the unique
crossing point as shown in Fig.~\ref{fig:345d} (b), (e), and (h) with 
$L^{\beta/\nu} m$ versus $\mu$ plotted, and then we present the
collapse of the numerical data into smooth curves in Fig.~\ref{fig:345d}
(c), (f), and (i) for 3D, 4D, and 5D, respectively. 
In the above finite-size scaling analysis, the critical
point $\mu_c$ and  the critical exponents are determined:
$\mu_c \approx 5.5$, $\beta \approx 0.33$, $\nu \approx 0.63$ in 3D,
$\mu_c \approx 5.25$, $\beta \approx 0.5$, $\nu \approx 0.5$ in 4D,
$\mu_c \approx 6.7$, $\beta \approx 0.5$, $\nu \approx 0.5$ in 5D.
Accordingly, we conclude that the DAP
belongs to the same universality class as for the equilibrium Ising model
in $d$-dimensions for $d=1,2,3,4$, and $5$~\cite{N.Goldenfeld} (see Table~\ref{tab:summary}).

\begin{figure}
\includegraphics[width=0.48\textwidth]{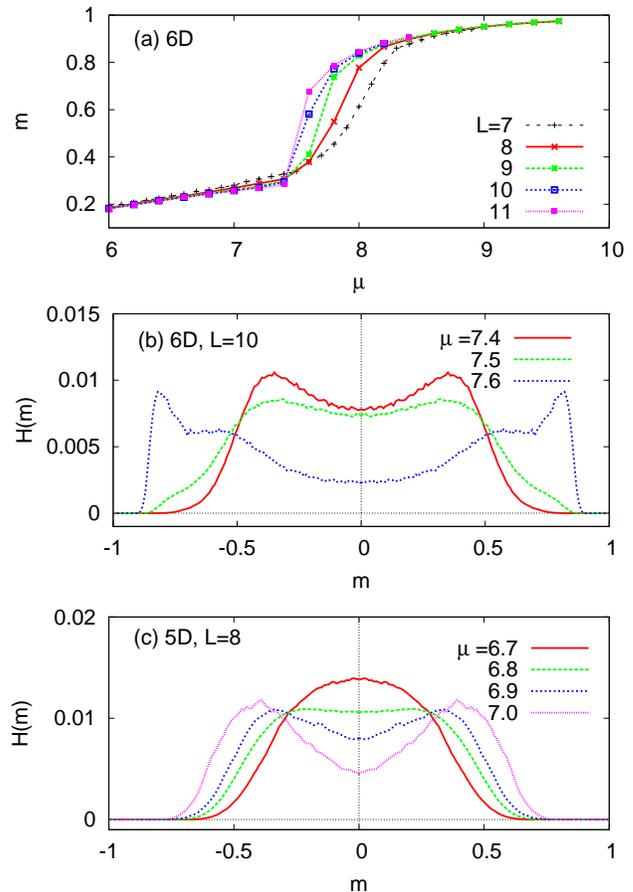}
\caption{(Color online) (a) $m$ versus $\mu$ in 6D. From the size dependence of
$m$, we roughly locate $\mu_c = 7.4(2)$.  (b) Normalized  histogram $H(m)$ of the
magnetization $m$ around the critical point for the system size $L=10$. The
peak position suddenly changes around $\mu_c(L=10)  \approx 7.5$, indicating
the discontinuous nature of the phase transition (compare with the histogram in
Ref.~\onlinecite{B.J.Kim2} for the discontinuous transition in the WS network).
At large values of $\mu$, the system often exhibits asymmetric $H(m)$. Only for
convenience of presentation, we symmetrized $H(m)$ to make it an even function of
$m$. For comparison, we also show in (c) $H(m)$ for the 5D regular lattice of the
size $L= 8$. Different from (b), the peak positions in $H(m)$ changes smoothly,
indicating the continuous nature of the phase transition, in accord with the
mean-field universality class revealed in Fig.~\ref{fig:345d}.
}
\label{fig:6d}
\end{figure}

\begin{table}[b]
\caption{Universality classes of the DAP
for $d$-dimensional regular lattices, for globally-coupled structure,
and for the WS network structure. The critical exponents $\beta$ and $\nu$
are included when the phase transition is of a continuous nature.
}
\begin{tabular}{ l | c | c |  c |  l  }
\hline
structure    & $\mu_c$ & $\beta$  & $\nu$ &  universality class       \\ \hline
1D regular   & $\infty$        &  $-$      &  $-$  &  1D Ising \\ 
2D regular~\cite{G.Szabo3,B.J.Kim2}   & 6.5    &  $1/8$    &  $1$  &  2D Ising  \\ 
3D regular  & 5.5         &  $0.33$    &  $0.63$  &  3D Ising  \\ 
4D regular  & 5.25         &  $1/2$    &  $1/2$  &  equilibrium mean-field  \\
5D regular  & 6.7          &  {$1/2$}    &  $1/2$  &  equilibrium mean-field  \\ 
6D regular  & 7.4         &  $-$    &  $-$  &  discontinuous \\
Global coupling~\cite{B.J.Kim2}   & $\infty$ &  $-$    &  $-$  &  no phase transition \\ 
WS$_1$ and WS$_2$   & $^{\rm *}$  &  $-$    &  $-$  &  discontinuous \\
\hline
\end{tabular}
\begin{tablenotes}
\item $^{\rm *}$ $\mu_c$ for WS$_1$ and WS$_2$ depends on the rewiring probability $\alpha$ (see Fig.~\ref{fig:wsphd}).
When $\alpha = 0$, WS$_1$ and WS$_2$ are identical to 1D and 2D regular lattices, respectively.
\end{tablenotes}
\label{tab:summary}
\end{table}

\subsection{DAP in 6-dimensional regular lattice}
\label{subsec:6d}
We next examine the nature of the phase transition in 6D.
Due to the practical limitation of the computational resources, we are limited
to use the linear size $L = 7, 8, 9, 10$, and 11.
Surprisingly, we find that the transition nature in 6D is very different from
the simple expectation of the equilibrium mean-field type
and becomes discontinuous, similarly to the WS$_1$
and WS$_2$ networks presented in Sec.~\ref{subsec:WS}.  
In Fig.~\ref{fig:6d}(a), the order parameter $m$ is shown as 
a function of $\mu$, exhibiting the transition
around $\mu_c \approx 7.4$. Similarly to the WS$_1$ and WS$_2$ networks,
the change of $m$ near $\mu_c$ becomes more abrupt as
$L$ is increased, which indicates the discontinuous nature of the 
phase transition. In Fig.~\ref{fig:6d}(b), we display the normalized histogram $H(m)$
of the order parameter $m$ [$m \equiv (c_0 + c_2 + c_4) - (c_1 + c_2 + c_3)$
has been used to plot $H(m)$], which clearly exhibits the signature of the discontinuous
phase transition, similarly to the WS network~\cite{B.J.Kim2}.
As another evidence of the non-mean-field nature of the transition in 6D, we also
plot $L^{\beta/\nu}$ as a function of $\mu$ in the same way as we did for 3D, 4D,
and 5D in Fig.~\ref{fig:345d}(b), (e), and (h)(not shown here). With the mean-filed values $\beta = \nu = 1/2$,
the curves do not make a unique crossing, which again supports the non-mean-field nature
of the phase transition in 6D.

\begin{figure}
\includegraphics[width=0.48\textwidth]{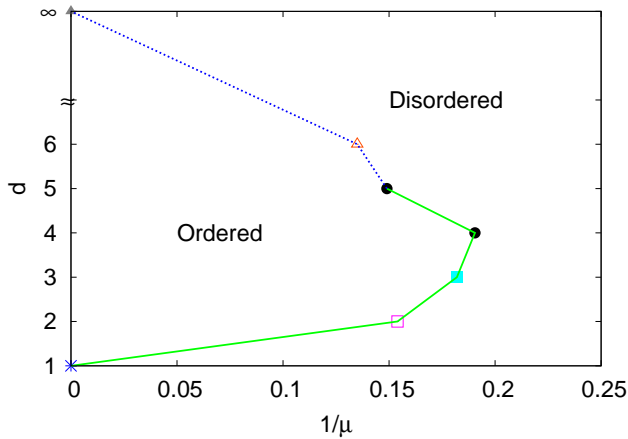}
\caption{\label{fig:ddphd}(Color online) Phase diagram for the DAP
in $d$-dimensional regular lattices with
the food web given in Fig.~\ref{fig:foodweb} in the plane
of the mutation parameter $1/\mu$ and $d$.
Different symbols denote different universality classes, and the
lines are only guides to eyes. For $d=1,2,3,4$, and 5, the DAP
belongs to the Ising universality class. As $d$ becomes
larger, the nature of phase
transition is changed to a discontinuous one.  The point at $d = \infty$
is included from the result in Ref.~\onlinecite{B.J.Kim2} for the globally-coupled
case. }
\end{figure}

In Fig.~\ref{fig:ddphd}, we summarize results for the phase transitions
of the DAP in  $d$-dimensional hypercubic regular lattices.
For $d\le 5$, the system shares the critical behavior with
the Ising model (see Table~\ref{tab:summary}): No phase transition in 1D, 
and continuous phase transitions in 2D, 3D, 4D, and 5D with critical
exponents corresponding to the Ising models in the same dimensions.
Very interestingly, as the dimensionality $d$ becomes
larger ($d=6$), the nature of the phase transition changes to a discontinuous
one. In a sharp contrast, the equilibrium Ising model in higher dimensions
than five belongs to the same mean-field universality class as in four 
dimension (up to logarithmic corrections),
which defines the upper critical dimension of the Ising model ($d_c = 4$).
On the other hand, the DAP displays
very different behavior: Although the equilibrium mean-field universality 
is identified in 4D and 5D, it does not lead to the conclusion that $d > 5$ should
exhibit the mean-field universality. 
From our extensive simulations of the DAP in the $d$-dimensional
hypercubic regular lattices, we propose that the system has three
different critical dimensions: (i) The usual lower critical dimension
$d_{\rm low} = 1$ below which the system is always disordered, (ii) the
first upper critical dimension $d_{\rm up}^{(1)} = 4$ splitting the non-mean-field
transition ($d < d_{\rm up}^{(1)}$)  and the mean-field transition 
($d \geq d_{\rm up}^{(1)}$). Different from the equilibrium
Ising model, the DAP model does not always show the mean-filed nature for 
dimensions higher than $d_{\rm up}^{(1)}$ and there exists (iii) 
the second upper critical dimension $d_{\rm up}^{(2)}=6$ beyond which 
the DAP shows the discontinuous phase transition.
Unfortunately, we do not
have any rigorous reasoning for or against the above ``conjecture.''

Although the nature of the phase transition cannot be determined 
by the mean-field theory in Ref.~\onlinecite{B.J.Kim2}, it 
predicts that the critical point $1/\mu_c$ in $d$ dimensions should approach to 0 from 
above as $d \rightarrow \infty$, because the globally-coupled 
network is equivalent to the infinite-dimensional systems. 
Consequently, the vanishing value $1/\mu_c = 0$  in both limits of $d \rightarrow \infty$
and $d=1$ makes us expect the reentrance behavior as $d$ is increased, 
which turns out to be true as shown in 
Fig.~\ref{fig:ddphd}. This is also very different
from the equilibrium Ising model in which the critical temperature
monotonically increases with the dimensionality.

\section{\label{sec:cluster}Cluster Mean-Field Calculation}
From our numerical observation that the discontinuous transition is not 
only due to the network property but it can also arise from the increased
dimensionality for hypercubic regular lattices, one would expect that 
there will be a mean-field theory to predict the discontinuous 
transition. Along this direction, we apply the cluster mean-field  (CMF)
theory~\cite{MD1999} to the two-dimensional model.
In this section, the lattice point in two-dimensional square lattice is denoted 
by $\bm{x}$ with the decomposition $\bm{x} = n_1 \bm{e}_1 + n_2 \bm{e}_2$, where
$n_i$'s are integers and $\bm{e}_i$ are unit vector along the direction $i$.
The species at site $\bm{x}$ will be denoted by $s_{\bm{x}}$. 

\begin{figure}
\includegraphics[width=0.48\textwidth]{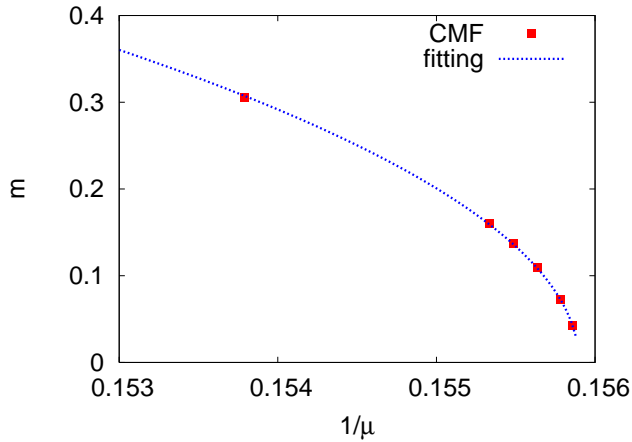}
\caption{\label{fig:cmf}(Color online) Magnetization $m$ as a function of $1/\mu$ 
in the cluster mean-field approximation. In the fitting, 
$m(\mu) = A (1/\mu_c - 1/\mu)^{1/2}$ is used with two fitting parameter $A$ and $\mu_c$,
resulting in $\mu_c \simeq 6.41$.}
\end{figure}

The approximation scheme will be detailed in the Appendix and here we just
sketch the calculation method. To begin with, we write down the exact evolution
equation for the marginal probability (see Appendix) $P_4(s_{\bm{x}},
s_{\bm{x}+\bm{e}_1},s_{\bm{x}+\bm{e}_2},s_{\bm{x}+\bm{e}_1+\bm{e}_2}) $ from
the master equation. Since each $s$ can take six different values
($s=0,1,\cdots, 5$) we have to find out $6^4$ equations. Thanks to the spatial
(rotation and inversion) and intra-alliance symmetry [$P(\{ s \} ) = P(\{s +2
\})$ with modulo 6], the number of equations is greatly reduced to 77, 
and the probability conservation condition reduces it to 76.
To get the steady state magnetization, we numerically iterate those 76
equations until the magnetization does not change significantly. The results
are summarized in Fig.~\ref{fig:cmf}. 
We then use the fitting function 
$m(\mu) \sim (1/\mu_c - 1/\mu)^{\beta}$ with the mean-field critical 
exponent $\beta = 1/2$ and estimate the critical point
$\mu_c \simeq 6.41$, which should be compared to the Monte Carlo simulation
results $\mu_c \simeq 6.50(4)$ in Table~\ref{tab:summary} and Ref.~\onlinecite{B.J.Kim2}. 
Unlike the higher-dimensional systems with $d \geq 6$, the CMF does not seem to predict the
discontinuous transition although it gives us an accurate estimation of the
critical point for the 2D system.  
However, CMF does not  provide a clue to why the models in 
higher dimensions above 5 as well as in WS networks
display the discontinuous transition. It is quite difficult to believe that 
the CMF with a larger cluster size than 2 can provide an explanation, 
because the actual transition in two dimensions is still continuous. 
One way might be applying the CMF to 6-dimensional systems, but it is simply 
too complicated to calculate.

\section{Summary}
\label{sec:summary}
We have investigated the alliance breaking
phase transitions of the six-species predator prey model called the
defensive alliance process (DAP) in 
the complex networks and also in the $d$-dimensional
regular lattices. Identified nature of the phase transition
is summarized in Table~\ref{tab:summary} with critical exponents
included when available.

In the WS network, we have observed that the alliance breaking
phase transition is of a discontinuous nature. 
Interesting reentrant phase transition in the WS network 
has been observed as the
rewiring probability $\alpha$ is increased at a fixed mutation rate,
implying the intricate role of the random shortcuts on the
phase transition.

Hypercubic regular lattice
structures in $d$ dimensions have also been used as underlying
spatial interaction structure of the DAP.
Observed is that the phase diagram in the plane of the dimensionality
and the mutation rate again displays an interesting reentrant
behavior (see Fig.~\ref{fig:ddphd}), which is in a striking contrast 
to the equilibrium Ising model. For the latter model, 
the critical temperature is simply an increasing function of the 
dimensionality. We have identified the universality class for various
dimensions ($d=1,2, \cdots, 6$) and $d=1, 2, \cdots, 5$ exhibits the
same critical behavior as for the equilibrium Ising model.
In contrast, as $d$ is increased further, discontinuous phase
transition has been observed for $d=6$.
In the hope to find a theory predicting the discontinuous transition, we 
have applied the cluster mean-field approximation for the two-dimensional systems,
but we only find the continuous transition within this scheme.
Still, the reason why the higher-dimensional systems exhibit the discontinuous
transition remains mystery, which can be an interesting theoretical question
to be pursued further.

\acknowledgments
This work was supported by the Korea Research Foundation 
funded by the Korean Government (MOEHRD) with the Grant No. KRF-2007-313-C00282.

\appendix*
\section{Two-dimensional cluster mean-field theory}
In this Appendix, the equations and symmetry of the cluster mean-field approximation
with the cluster size two will be detailed. 
We start from deriving the exact equation for the marginal probability for the
local configuration at four sites $\bm{x}$, $\bm{x} + \bm{e}_1$, $\bm{x}+\bm{e}_2$, and 
$\bm{x}+\bm{e}_1+\bm{e}_2$ ($\bm{x}$ is the vector designating a lattice point and 
$\bm{e}_i$'s are unit vectors along $i$-th direction). 

The marginal probability $P_4$ is defined by
\begin{widetext}
\begin{equation}
P_4(a,b,c,d) \equiv P_4 \begin{pmatrix}a&b\\c&d\end{pmatrix}\equiv 
\sum{}^\prime P\left (\ldots,s_{\bm{x}} = c,s_{\bm{x}+\bm{e}_1 }=d,
s_{\bm{x}+\bm{e}_2} = a, s_{\bm{x}+\bm{e}_1+\bm{e}_2} = b,\ldots
\right ),
\end{equation}
\end{widetext}
where the time dependence is implicitly assumed and 
the primed-sum ($\sum^\prime$) run over the all possible configurations with four specified sites having
the denoted species ($0\le a,b,c,d \le 5$).
We always assume the modulo-6 equivalence among species indices, e.g., 
species 6 is equal to species 0, and so on.
Due to the translational invariance, $P_4$ does not depend on $\bm{x}$ if 
initial condition does not have $\bm{x}$ dependence or if we are only 
interested in the stationary state. Hence we can safely omit the explicit $\bm{x}$ 
dependence for the function $P_4$.

\begin{widetext}
The exact time evolution for $P_4$ can be written as 
\begin{eqnarray}
\frac{d}{dt} \fPC 
= \frac{P}{2} \sum_{i=1}^2 \left [ 
\fPp{1} +
\fPp{2} +
\fPp{3} +
\fPp{4} \right ]- 4P \fPC 
 \nonumber\\
+ \frac{1-P}{2} \left \{ \sum_{i=1}^2 
\left [ \Pnwup + \Pnwle + \Pneup + \Pneri + 
 \Pseri  \right . \right . \nonumber \\  + \Psedo\left .+\Pswdo + \Pswle \right ]%\nonumber \\
-  \sum_{i=4}^5 \Bigg [ \mPnwup + \mPnwle \nonumber \\ + \mPneup + \mPneri +
 \mPseri + \mPsedo +  \mPswdo \nonumber \\+ \mPswle \Bigg ] 
+\sum_{i=1}^2  \Bigg [ (\delta_{ab} + \delta_{ac}) \fPp{1}
+(\delta_{ab} + \delta_{bd}) \fPp{2}
+(\delta_{ac} + \delta_{cd}) \fPp{3}\nonumber\\ 
+(\delta_{bd} + \delta_{cd}) \fPp{4} \Bigg ]
- \left . 
\sum_{i=1}^5 (1-\delta_{i3})(\delta_{a,b+i} 
+ \delta_{a,c+i} + \delta_{b,d+i} + \delta_{c,d+i} ) \fPC 
\right \}.
\label{Equation:mean_field}
\end{eqnarray}
\end{widetext}
In Eq.~\eqref{Equation:mean_field}, we also introduced 
$P_5$ which is the marginal probability of the local configuration with 
5 sites, taking the form given as an argument. Due to the rotational and mirror symmetry, 
one can easily see that
\begin{equation}
P_5\begin{pmatrix}a&b\\c&d\\e&\end{pmatrix} 
= P_5\begin{pmatrix}e&c&a\\&d&b\end{pmatrix}
= P_5\begin{pmatrix}&d&b\\e&c&a\end{pmatrix}.
\label{Equation:P5_symmetry}
\end{equation}
Hence we do not have to deal with 8 different local configurations independently 
while treating $P_5$. 

Due to the hierarchy appearing in Eq.~\eqref{Equation:mean_field} ($P_5$ is not 
reducible in terms of $P_4$),
it is not easy to solve it exactly, which necessitates the use of the approximation scheme.
To this end, we treat the correlation within squares of linear size 2 completely
and neglect the correlation beyond linear length 2. To be specific, we are using the
approximation scheme such that~\cite{MD1999}
\begin{equation}
P_5\begin{pmatrix}a&b\\c&d\\e&\end{pmatrix} \simeq 
\frac{\fPC P_3\begin{pmatrix}c&d\\e&\end{pmatrix}}{P_2\begin{pmatrix}c&d\end{pmatrix}},
\end{equation}
where $P_2$ and $P_3$ are marginal probabilities defined similarly to $P_4$. 
The probability conservation makes
it possible to find $P_2$ and $P_3$ once we know $P_4$ from the relations
\begin{eqnarray}
P_3\begin{pmatrix}c&d\\e&\end{pmatrix} = \sum_{i=0}^5
P_4\begin{pmatrix}c&d\\e&i\end{pmatrix},\\
P_2\begin{pmatrix}c&d\end{pmatrix} = \sum_{i=0}^5
P_3\begin{pmatrix}c&d\\i&\end{pmatrix}.
\end{eqnarray}
The above approximation scheme along with the symmetry consideration given
in Eq.~\eqref{Equation:P5_symmetry} lets
Eq.~\eqref{Equation:mean_field} closed in the sense that the number of equations are
equal to that of variables. Since it is still infeasible to solve the approximate
equation analytically, we resort to the numerical solutions.

One may treat the $6^4 = 1296$ equations directly without considering the degeneracy
of $P_4$'s. However, symmetry consideration reduces the number of equations we have to
deal with considerably. Now we will show that actually we have only to treat 76 
equations to get the full solution.

There are three symmetry operations which are summarized as follows:
\begin{eqnarray}
P_4\begin{pmatrix}a&b\\c&d\end{pmatrix} = P_4\begin{pmatrix}b&d\\a&c\end{pmatrix}
%= P_4\begin{pmatrix}d&c\\b&a\end{pmatrix}
=P_4\begin{pmatrix}c&a\\d&b\end{pmatrix},\\
P_4\begin{pmatrix}a&b\\c&d\end{pmatrix}=P_4\begin{pmatrix}
b&a\\d&c\end{pmatrix}
%= P_4\begin{pmatrix}c&d\\a&b\end{pmatrix}
=P_4\begin{pmatrix}a&c\\b&d\end{pmatrix}
%= P_4\begin{pmatrix}d&b\\c&a\end{pmatrix}
,\\
P_4\begin{pmatrix}a&b\\c&d\end{pmatrix}
=P_4\begin{pmatrix}a+2&b+2\\c+2&d+2\end{pmatrix},
\end{eqnarray}
which are rotation ($\pm 90^\circ$), 
mirror (against the diagonal axes), and intra-ally symmetric operations, respectively.
To find the degeneracy due to symmetry, 
let us first categorize the local configurations according to the number of same species
among 4 sites. There are 5 such categories which take the form $\{0,0,0,0\}$ (6), 
$\{0,0,0,1\}$ (120), $\{0,0,1,1\}$ (90), $\{0,0,1,2\}$ (720), and $\{0,1,2,3\}$ (360), 
where the species numbers in curly brackets are just for the representative purpose, 
the order of four elements in the curly braces are irrelevant, and 
the numbers in parentheses indicate the total number of local
configurations of the corresponding categories whose sum should be $6^4$.

By considering the above symmetry operations, we reduce the number of independent
variables. When $a=b=c=d$, the intra-ally symmetry reduces the independent
variables from 6 to 2.  When three of four species are the same such as $a=b=c\neq d$,
rotational symmetry as well as the intra-ally one reduces the number of 
independent variables.
For example, $P_4(a,a,a,d) = P_4(a,a,d,a) = P_4(a,d,a,a) = P_4(d,a,a,a)$ by rotation
and $P_4(a,a,a,d)  = P_4(a+2,a+2,a+2,d+2)$ by intra-ally transformations.
From this consideration, one can find that there are 10 independent variables
out of 120 variables. In case $a=b\neq c=d$, the symmetry consideration shows that
(we omit $P_4$ for convenience)
\begin{eqnarray}
\begin{pmatrix}a&a\\c&c\end{pmatrix}=
\begin{pmatrix}a&c\\a&c\end{pmatrix}=
\begin{pmatrix}c&c\\a&a\end{pmatrix}=
\begin{pmatrix}c&a\\c&a\end{pmatrix},\nonumber\\
\begin{pmatrix}a&c\\c&a\end{pmatrix}=
\begin{pmatrix}c&a\\a&c\end{pmatrix},
\end{eqnarray}
which reduces the independent variables from 90 to 30.
The intra-ally symmetry once again reduces the number from 30
to 10.

The fourth case is $a=b\neq c\neq d\neq a$.
The symmetry enforces the equivalence among configurations such that
\begin{eqnarray}
\begin{pmatrix}a&a\\c&d\end{pmatrix}=
\begin{pmatrix}a&d\\a&c\end{pmatrix}=
\begin{pmatrix}d&c\\a&a\end{pmatrix}=
\begin{pmatrix}c&a\\d&a\end{pmatrix},\nonumber\\
\begin{pmatrix}a&c\\d&a\end{pmatrix}=
\begin{pmatrix}c&a\\a&d\end{pmatrix},
\label{Equation:aacd}
\end{eqnarray}
where exchanging $c$ with $d$, which is equivalent to the
mirror transformation, also gives the equivalent configurations.
Hence Eq.~\eqref{Equation:aacd} combined with the intra-ally symmetry
reduces the number from 720 to 40.
The last category contains all different species. This category has $\binom{6}{4}=15$ 
elements. Since the rotation and the mirror symmetry operations always conserve the diagonal 
relations, each set has 3 different classes. Due to the intra-ally symmetry, however,
only 15 configurations remain independent. 
To sum all independent configurations, we find that
there are 77 (=2+10+10+40+15) 
independent variables. Due to the probability conservation, one equations
becomes reducible by other variables. So the final number of independent variables is 76. 

One may solve the stationary state solution by setting $\partial/\partial t = 0$ in
Eq.~\eqref{Equation:mean_field}. For us, this turned out not to be easy, so we
numerically integrate Eq.~\eqref{Equation:mean_field} starting from the fully
magnetized state with intra-ally symmetry such that $P_4(a,b,c,d) = 1/3^n$ with
$n$ to be the number of members in alliance I among $a,b,c,d$ and found the
stationary state magnetization, which is summarized in Fig.~\ref{fig:cmf}.

\end{document}